\documentstyle[12pt]{article}
\advance \textheight by \topskip
\makeatletter
\def\eqnarray{\stepcounter{equation}\let\@currentlabel=\theequation
\global\@eqnswtrue
\global\@eqcnt\z@\tabskip\@centering\let\\=\@eqncr
$$\halign to \displaywidth\bgroup\@eqnsel\hskip\@centering
  $\displaystyle\tabskip\z@{##}$&\global\@eqcnt\@ne
  \hfil$\displaystyle{\hbox{}##\hbox{}}$\hfil
  &\global\@eqcnt\tw@ $\displaystyle\tabskip\z@
  {##}$\hfil\tabskip\@centering&\llap{##}\tabskip\z@\cr}
\@addtoreset{equation}{section}
  \def\theequation{\thesection.\arabic{equation}}
\makeatother
\mathchardef\Pi="105
\mathchardef\Phi="108
\def\mbar#1{\kern 0.1em \overline{\kern -0.1em #1\kern -0.1em} \kern 0.1em}
\def\vp{\varphi}

\begin{document}

\begin{titlepage}
\hbox to \hsize{\hfil hep-th/9307161}
\hbox to \hsize{\hfil IHEP 92--146}
\hbox to \hsize{\hfil October, 1992}
\vfill
\large \bf
\begin{center}
THE CANONICAL SYMMETRY\\
FOR INTEGRABLE SYSTEMS
\end{center}
\vskip 1cm
\normalsize
\begin{center}
{\bf A. N. Leznov and A. V. Razumov\footnote{E--mail:
razumov@mx.ihep.su}}\\
{\small Institute for High Energy Physics, 142284 Protvino, Moscow Region,
Russia}
\end{center}
\vskip 2.cm
\begin{abstract}
\noindent
The properties of discrete nonlinear symmetries of integrable
equations are investigated. These symmetries are shown to be
canonical transformations. On the basis of the considered examples, it is
concluded, that the densities of the conservation laws are changed under
these transformations by spatial divergencies.
\end{abstract}
\vfill
\end{titlepage}

\section{Introduction}

Recently the discrete nonlinear symmetry transformations for the variety
of the integrable systems were obtained.\cite{Lez91a,Lez91b,Lez92} These
transformations were given in an explicit form, which allows one to get
the whole hierarchy of solutions from any initial solution of the given
nonlinear equation. In particular, the soliton solutions can be obtained
with the help of these transformations without using the inverse
scattering method. Unfortunately, we do not know a constructive method to
find the corresponding discrete symmetry transformation for an arbitrary
integrable systems. Actually such transformations arose as a byproduct of
the algebraic approach to the construction of the soliton solutions of
integrable systems.\cite{LSa}

In this paper we investigate the properties of the discrete symmetry from
the point of view of the Hamiltonian structure relevant to integrable
systems. Numerous examples, not only two dimensional ones, lead us to the
conclusion that all these transformations are canonical.

For the case of the nonlinear Schr\"odinger equation, the densities of the
lowest conservation laws appeared to be shifted under these
transformations by total space derivatives. It can be shown that all the
densities of the conservation laws possess this property, and, in a sense,
these densities are invariants of the transformations under consideration.
Because of the importance of this property we shall devote to its proof a
separate paper.

Note that the symmetry transformations, considered in the present paper,
are very different from the well known discrete symmetry transformations
of integrable systems, called B\"acklund transformations.\cite{Back} The
connection between those transformations and our ones is presently
unclear. Thus, we have found it useful to call our transformations
differently, and the results described below have prompted us to use for
them the term `canonical symmetry transformations'.  Nevertheless, we
would like to point out that the usual B\"acklund transformation are also
canonical transformations (see, for example, Ref.~7).

\section{Nonlinear Schr\"odinger Equation}

As the first example we consider the canonical symmetry of
the nonlinear Schr\"o\-din\-ger equation. It is a nonlinear equation for a
complex valued function $\psi(x, t)$ of the form
\begin{equation}
i \dot \psi + \psi'' - 2\kappa |\psi|^2 \psi = 0. \label{2.1}
\end{equation}
Here and below dot means the partial derivative over $t$ and prime --- the
partial derivative over $x$. The real parameter $\kappa$ is the coupling
constant. We denote the complex conjugation by bar, so that $|\psi|^2 =
\mbar \psi \psi$. Note, that we also have the complex conjugate
equation
\begin{equation}
i \dot {\mbar \psi} - \mbar \psi'' + 2 \kappa |\psi|^2 \mbar
\psi = 0.
\label{2.2}
\end{equation}
Equations (\ref{2.1}) and (\ref{2.2}) are the Euler--Lagrange equations
for the La\-gran\-gian
\begin{equation}
L = \int dx\, \left(\frac{i}{2} (\mbar \psi \dot \psi - \dot {\mbar \psi}
\psi) - |\psi'|^2 - \kappa |\psi|^4 \right). \label{2.4}
\end{equation}

We see that the action, corresponding to the Lagrangian (\ref{2.4}), is
already in the Hamiltonian form. Thus we can consider the space with the
coordinates $\psi(x)$ and $\mbar \psi(x)$ as the phase space of the
system. The nonzero Poisson brackets in this case are
\begin{equation}
\{\psi(x),\, \mbar \psi(x')\} = -i\delta(x - x'). \label{2.5}
\end{equation}
These Poisson brackets are generated by the symplectic 2--form
\begin{equation}
\Omega = i \int dx\, d\mbar \psi \wedge d\psi. \label{2.6}
\end{equation}
For the Hamiltonian of the system we have the expression
\begin{equation}
H = \int dx\, (|\psi'|^2 + \kappa |\psi|^4). \label{2.7}
\end{equation}
It is easy to verify that this Hamiltonian with the Poisson brackets,
defined by (\ref{2.5}), leads to nonlinear Schr\"odinger equation
(\ref{2.1}), (\ref{2.2}).

To define the canonical symmetry transformation we should construct a
complex extension of the system. To this end we denote $q = \psi$, $r =
\mbar \psi$ and consider $q$ and $r$ as independent complex valued
functions, satisfying the following system of equations
\begin{equation}
i\dot q + q'' - 2 \kappa r q^2 = 0, \qquad i\dot r - r'' + 2 \kappa q r^2
= 0. \label{2.13}
\end{equation}
These equations can be derived from the action with the density of the
Lagrangian
\begin{equation}
{\cal L} = \frac{i}{2} (r \dot q - \dot r q) - q'r' - \kappa q^2 r^2.
\label{2.14}
\end{equation}
This density of the Lagrangian is complex valued and not appropriate
to formulate the Hamiltonian description of the system. We will use
a real Lagrangian
\begin{equation}
L = \frac{1}{2} \int dx\, ({\cal L} + \mbar{\cal L}), \label{2.15}
\end{equation}
which leads to the same equations.

Proceeding as above, we get the phase space with the coordinates $q$, $r$
and $\mbar q$, $\mbar r$. The symplectic 2--form in this case is of the
form
\begin{equation}
\Omega = \frac{i}{2} \int dx\, (dr \wedge dq - d\mbar r \wedge d\mbar q),
\label{2.16}
\end{equation}
and for the nonzero Poisson brackets we have the expressions
\begin{equation}
\{q(x), r(x')\} = -2i \delta(x - x'), \qquad \{\mbar q(x), \mbar
r(x')\} = 2i \delta(x - x'). \label{2.17}
\end{equation}
The Hamiltonian of the system is
\begin{equation}
H = \frac{1}{2} \int dx\, ({\cal H} + \mbar {\cal H}), \label{2.18}
\end{equation}
where
\begin{equation}
{\cal H} = r'q' + \kappa r^2 q^2. \label{2.19}
\end{equation}
Below, considering complex extensions, we use Lagrangians and Hamiltonians
of form (\ref{2.15}) and (\ref{2.18}).  In such cases we call ${\cal
L}$ and ${\cal H}$ the density of the Lagrangian and the Hamiltonian,
respectively.

To get the initial nonlinear Schr\"odinger system it is enough to perform
the reduction to the surface, given by the equations
\begin{equation}
r = \mbar q, \qquad \mbar r = q. \label{2.20}
\end{equation}

The canonical symmetry transformation in this case has the
form\cite{Lez91a,Lez92,SYa91}
\begin{equation}
Q = \frac{1}{\kappa r}, \qquad R = \kappa r^2 q - r'' + \frac{r'^2}{r}.
\label{2.21}
\end{equation}
There are of course the corresponding formulae for the complex conjugated
functions $\mbar Q$ and $\mbar R$.  Usually we will not write such
formulae, because they can be obtained simply by the complex conjugation.
Note, that the surface, given by (\ref{2.20}), is not invariant under
transformation (\ref{2.21}), and it is the reason for introducing a
complex extension of the system.

We are going to show that transformation (\ref{2.21}) is canonical. To do
this it is convenient to use the method of generating functions
(functionals).\cite{Gol50} In our case this method is as follows.

The symplectic 2--form $\Omega$, given by (\ref{2.16}) is exact and can be
represented as
\begin{equation}
\Omega = d \Theta,  \label{2.22}
\end{equation}
where
\begin{equation}
\Theta = \frac{i}{2} \int dx\, (r dq - \mbar r d\mbar q). \label{2.23}
\end{equation}
Suppose that transformation from the variables $q$, $r$ and $\mbar q$,
$\mbar r$ to the variables $Q$, $R$ and $\mbar Q$, $\mbar R$ is canonical.
In this case we must have
\begin{equation}
\frac{i}{2} \int dx\, (r dq - \mbar r d \mbar q) = \frac{i}{2} \int dx\,
(R dQ - \mbar R d \mbar Q) + dF, \label{2.24}
\end{equation}
where $F$ is some real valued functional on the phase space. Suppose
further that the transformation under consideration is such that the
functions $q$, $Q$ and $\mbar q$ and $\mbar Q$ can be considered as
coordinates on the phase space. Representing $dF$ in the form
\begin{equation}
dF = \int dx\, \left( \frac{\delta F}{\delta q} dq + \frac{\delta
F}{\delta Q} dQ + \frac{\delta F}{\delta \mbar q}
d \mbar q + \frac{\delta F}{\delta \mbar Q} d \mbar Q \right),
\label{2.25}
\end{equation}
we get the following formulae for $r(x)$ and $R(x)$:
\begin{equation}
r = - 2i \frac{\delta F}{\delta q}, \qquad R = 2i \frac{\delta F}{\delta
Q}. \label{2.26}
\end{equation}
It is also clear that for any real valued functional $F$ of the variables
$q$, $Q$, and $\mbar q$, $\mbar Q$ formulae (\ref{2.26}) define a
canonical transformation, provided that they can be written as the
relations, defining an invertible transformation. The functional $F$ is
called the generating functional of the corresponding transformation.

Let us return again to our example. From (\ref{2.21}) we get
\begin{equation}
r = \frac{1}{\kappa Q}, \qquad R = \frac{1}{\kappa Q^2} \left(q + Q'' -
\frac{Q'^2}{Q} \right). \label{2.27}
\end{equation}
It is clear that we can consider the functions $q$, $Q$ and $\mbar q$,
$\mbar Q$ as coordinates on the phase space. Using now (\ref{2.27}) in
(\ref{2.26}), we get the following expression for the generating
functional:
\begin{equation}
F = \frac{1}{2} \int dx\, ({\cal F} + \mbar {\cal F}), \label{2.28}
\end{equation}
where
\begin{equation}
{\cal F} = \frac{i}{\kappa} \left( \frac{q}{Q} + \frac{1}{2}
\frac{Q'^2}{Q^2} \right). \label{2.29}
\end{equation}
Thus, we see that in this case the canonical symmetry is a
canonical transformation.

Present now the formula that describes the behaviour of the density of
the Hamiltonian ${\cal H}$ under the canonical symmetry transformation:
\begin{equation}
{\cal H}(Q,R) = {\cal H}(q,r) + \left(-2 q r' + \frac{r'' r'}{\kappa r^2}
- \frac{2}{3} \frac{r'^3}{\kappa r^3}\right)'. \label{2.30}
\end{equation}
{}From this relation we conclude, that under appropriate boundary conditions
the Hamiltonian of the system is invariant under the canonical symmetry
transformation.

The nonlinear Schr\"odinger equation is known to have an infinite set of
local conservation laws.\cite{ZSh71,FTa87} The same is true for the complex
extension of the system, which is considered here. The densities of the
lowest conservation laws are
\begin{eqnarray}
{\cal P}_1 &=& rq, \qquad {\cal P}_2 = -irq', \label{2.31} \\
{\cal P}_3 &=& -rq'' + \kappa r^2 q^2, \label{2.32} \\
{\cal P}_4 &=& i (rq''' - \kappa rq (r'q + 4 rq')). \label{2.33}
\end{eqnarray}
Through direct calculation we can show that these densities are invariant
under the canonical symmetry transformation up to the total derivatives,
i.\ e.
\begin{equation}
{\cal P}_n(Q,R) = {\cal P}_n(q,r) + {\cal D}'_n(q,r), \label{2.34}
\end{equation}
where
\begin{eqnarray}
{\cal D}_1 &=& -\frac{1}{\kappa} \frac{r'}{r}, \qquad {\cal D}_2 =
i\left(rq - \frac{1}{2\kappa} \frac{r'^2}{r^2} \right), \label{2.35} \\
{\cal D}_3 &=& rq' - r'q + \frac{1}{3\kappa} \frac{r'^3}{r^3},
\label{2.36} \\
{\cal D}_4 &=& i\left( - rq'' + \frac{3}{2} \kappa r^2 q^2 + r'q' -
\frac{r'^2q}{r} + \frac{1}{4\kappa} \frac{r'^4}{r^4} \right). \label{2.37}
\end{eqnarray}
It can be shown that all the densities of the local conservation laws obey
this property. We shall give the proof of this fact in the subsequent
paper. This property means, in particular, that the higher nonlinear
Schr\"odinger equations are invariant under canonical symmetry
transformation (\ref{2.21}).

\section{Modified Nonlinear Schr\"odinger Equation}

The modified nonlinear Schr\"odinger equation is the equation of the
form\cite{CLL79}
\begin{equation}
i \dot \psi + \psi'' - 2i \kappa |\psi|^2 \psi' = 0. \label{3.1}
\end{equation}
The corresponding complex extension is the following system of equations:
\begin{equation}
i \dot q + q'' - 2i \kappa (rq) q' = 0, \qquad i \dot r - r'' - 2i \kappa
(rq) r' = 0. \label{3.2}
\end{equation}
We consider this system as Euler--Lagrange equations, corresponding to the
Lagrangian
\begin{equation}
L = \frac{1}{2} \int dx\, ({\cal L} + \mbar{\cal L}), \label{3.3}
\end{equation}
where
\begin{equation}
{\cal L} = \frac{i}{2} (r\dot q - \dot r q) - r'q' - \frac{i\kappa}{2}
(r^2 qq' - q^2 rr'). \label{3.4}
\end{equation}

{}From (\ref{3.4}) we conclude that we can consider the space with the
coordinates $q$, $r$ and $\mbar q$, $\mbar r$ as the phase space of the
system, with the nonzero Poisson brackets (\ref{2.17}). The density of the
Hamiltonian is given by
\begin{equation}
{\cal H} = q'r' + \frac{i\kappa}{2}(r^2 qq' - q^2 rr'). \label{3.5}
\end{equation}

The canonical symmetry transformation for the case under consideration
is\cite{Lez91a,Lez92}
\begin{equation}
Q = \frac{1}{\kappa r}, \qquad R = \kappa r^2 q + i \left(r' -
\frac{r''r}{r'}\right). \label{3.6}
\end{equation}
Writing this transformation in the form
\begin{equation}
r = \frac{1}{\kappa Q}, \qquad R = \frac{1}{\kappa Q^2} \left[q + i
\left(Q' - \frac{Q''Q}{Q'}\right)\right], \label{3.7}
\end{equation}
and using (\ref{2.26}), we can show that canonical symmetry
transformation (\ref{3.6}) is a canonical transformation with the
generating functional, defined by the density
\begin{equation}
{\cal F} = \frac{i}{\kappa} \left(\frac{q}{Q} + i \frac{Q'}{Q} \ln Q'
\right). \label{3.8}
\end{equation}

The formula that describes the behaviour of the density of the Hamiltonian
(\ref{3.5}) under canonical symmetry transformation (\ref{3.6}) has the form
\begin{equation}
{\cal H}(Q,R) = {\cal H}(q,r) + \frac{1}{2} \left[-3 r'q - i\kappa r^2 q^2
- \frac{r''rq}{r'} + \frac{i}{\kappa} \left(\frac{r''}{r} - \frac{1}{2}
\frac{r'^2}{r^2} + \frac{1}{2} \frac{r''^2}{r'^2} \right) \right]'.
\end{equation}

\section{Derivative Nonlinear Schr\"odinger Equation}

The derivative nonlinear Schr\"odinger equation has the form\cite{KNe78}
\begin{equation}
i \dot \psi + \psi'' - 2i\kappa (|\psi|^2 \psi)' = 0. \label{4.1}
\end{equation}
Thus, we have the following complex extension
\begin{equation}
i \dot q + q'' - 2i\kappa (rq^2)' = 0, \qquad i \dot r - r'' - 2i\kappa
(r^2 q)' = 0. \label{4.2}
\end{equation}

Let us give at once the Hamiltonian formulation for this system. The phase
space is again the space with the coordinates $q$, $r$ and $\mbar q$,
$\mbar r$, but the nonzero Poisson brackets have now the form
\begin{equation}
\{q(x),\, r(x')\} = 2\delta' (x - x'), \qquad \{\mbar q(x),\, \mbar r(x')\}
= 2\delta' (x - x'), \label{4.3}
\end{equation}
and the density of the Hamiltonian is
\begin{equation}
{\cal H} = \frac{i}{2}(rq' - r'q) + \kappa r^2 q^2.
\end{equation}

Poisson brackets (\ref{4.3}) are generated by the symplectic 2--form
\begin{equation}
\Omega = \frac{1}{2} \int dx\, (\partial^{-1} dr \wedge dq +
\partial^{-1} d \mbar r \wedge d \mbar q), \label{4.4}
\end{equation}
where the operator $\partial^{-1}$ acts on a function $f(x)$, for example,
according to the rule
\begin{equation}
\partial^{-1} f(x) = \frac{1}{2} \left(\int_{-\infty}^x dx'\, f(x') -
\int_x^{\infty} dx'\, f(x') \right). \label{4.5}
\end{equation}
The 2--form $\Omega$ is exact and can be represented as
\begin{equation}
\Omega = d\Theta, \label{4.6}
\end{equation}
where
\begin{equation}
\Theta = \frac{1}{2} \int dx\, (\partial^{-1} r dq + \partial^{-1}
\mbar r d\mbar q). \label{4.7}
\end{equation}

The canonical symmetry transformation for the derivative nonlinear
Schr\"odinger equation is of the form\cite{Lez91a,Lez92}
\begin{equation}
Q = r, \qquad R = q - \frac{i}{\kappa} \frac{r'}{r^2}. \label{4.8}
\end{equation}
As we have a noncanonical symplectic 2--form, in this case we can not use
formulae (\ref{2.26}). Using the method, considered in section 2, we get
that now the corresponding formulae are
\begin{equation}
r = 2 \left( \frac{\delta F}{\delta q} \right)', \qquad
R = - 2 \left( \frac{\delta F}{\delta Q} \right)'.
\label{4.9}
\end{equation}
Writing now the canonical symmetry transformation as
\begin{equation}
r = Q, \qquad R = q - \frac{i}{\kappa} \frac{Q'}{Q^2}, \label{4.10}
\end{equation}
and using (\ref{4.9}), we see that canonical symmetry transformation
(\ref{4.8}) is canonical and the density of the generating functional is
\begin{equation}
{\cal F} = q \partial^{-1} Q - \frac{i}{\kappa} \ln Q. \label{4.11}
\end{equation}

The behaviour of the quantity ${\cal H}$ under the canonical symmetry
transformation is described by the formula
\begin{equation}
{\cal H}(Q, R) = {\cal H}(q,r) + \left( -iqr - \frac{1}{2\kappa}
\frac{r'}{r}\right)'. \label{4.12}
\end{equation}

\section{Heisenberg Model}

The phase space of the Heisenberg model is formed by real vector functions
$\vec S(x) = (S_1(x), S_2(x), S_3(x))$, subjected to the condition
\begin{equation}
\vec S^2(x) = 1. \label{7.1}
\end{equation}
The equations of motion have the form
\begin{equation}
\dot{\vec S} =  \vec S \times \vec S''. \label{7.2}
\end{equation}
The Poisson brackets are given by
\begin{equation}
\{S_i(x), S_j(x')\} = - \varepsilon _{ijk} S_k(x) \delta(x - x').
\label{7.3}
\end{equation}
It is easy to show that equations (\ref{7.1}) can be written
as the Hamilton equations
\begin{equation}
\dot{\vec S} = \{\vec S, H\}, \label{7.4}
\end{equation}
with the Hamiltonian
\begin{equation}
H = \frac{1}{2} \int dx\, (\vec S')^2. \label{7.5}
\end{equation}

Introduce the stereographic coordinates
\begin{equation}
\psi = \frac{S_1 + iS_2}{1 + S_3}, \qquad \mbar \psi = \frac{S_1 - iS_2}{1
+ S_3}. \label{7.6}
\end{equation}
For the Poisson brackets of $\psi$ and $\mbar\psi$ we find the expressions
\begin{eqnarray}
&\{\psi (x), \psi (x')\} = 0, \qquad \{\mbar \psi(x), \mbar \psi(x')\} =
0,& \label{7.8} \\
&\{\psi (x), \mbar \psi (x')\} = - \displaystyle \frac{i}{2} (1 +
|\psi|^2(x))^2 \delta(x - x'),& \label{7.9}
\end{eqnarray}
and the Hamiltonian in terms of the new variables becomes
\begin{equation}
H = \int dx\, \frac{2 |\psi'|^2}{(1 + |\psi|^2)^2}. \label{7.10}
\end{equation}
Using these relations, we see that the equations of motion can be written
in the form
\begin{equation}
i\dot \psi + \psi'' - \frac{2\mbar \psi \psi'^2}{1 + |\psi|^2} = 0.
\label{7.11}
\end{equation}

Consider as a complex extension the following system of equations
\begin{equation}
i \dot q + q'' - \frac{2 r q'^2}{1 + rq} = 0, \qquad i \dot r - r'' +
\frac{2 q r'^2}{1 + rq} = 0. \label{7.12}
\end{equation}
In fact, we can get this complex extension, if from the beginning suppose
that $\vec S$ is a complex valued vector function. The nonzero Poisson
brackets in this case are
\begin{eqnarray}
&\{q(x), r(x')\} = -i (1 + rq(x))^2 \delta(x - x'),& \label{7.13a} \\
&\{\mbar q(x), \mbar r(x')\} = i (1 + \mbar r \mbar q(x))^2 \delta(x -
x'),& \label{7.13b}
\end{eqnarray}
and the density of the Hamiltonian is given by
\begin{equation}
{\cal H} = \frac{2 r'q'}{(1 + rq)^2}. \label{7.14}
\end{equation}

Poisson brackets (\ref{7.13a}), (\ref{7.13b}) are generated by the
symplectic  2--form
\begin{equation}
\Omega = i \int dx\, \left( \frac{dr \wedge dq}{(1 + rq)^2} - \frac{d\mbar
r \wedge d\mbar q}{(1 + \mbar r \mbar q)^2}\right), \label{7.15}
\end{equation}
which can be represented as
\begin{equation}
\Omega = d\Theta, \label{7.16}
\end{equation}
where
\begin{equation}
\Theta = -i \int dx\, \left( \frac{dq}{q(1 + rq)} - \frac{d\mbar q}{\mbar
q(1 + \mbar r\mbar q)} \right). \label{7.17}
\end{equation}

Consider now a canonical transformation from the variables $q$, $r$ and
$\mbar r$, $\mbar q$, to the variables $Q$, $R$ and $\mbar Q$, $\mbar R$,
and suppose that the variables $q$, $Q$ and $\mbar q$, $\mbar Q$ can be
considered as coordinates on the phase space. Using (\ref{7.17}), it can be
directly shown that in this case we can write
\begin{equation}
\frac{1}{q(1 + rq)} = i \frac{\delta F}{\delta q}, \qquad \frac{1}{Q(1 +
RQ)} = -i \frac{\delta F}{\delta Q}, \label{7.18}
\end{equation}
where $F$ is a functional of $q$, $Q$ and $\mbar q$, $\mbar Q$.

The canonical symmetry transformation in this case has the form
\begin{equation}
Q = \frac{1}{r}, \qquad \frac{1}{1 + RQ} - \frac{1}{1 + rq} =
\frac{r r'' - r'^2}{r'^2}. \label{7.19}
\end{equation}
{}From (\ref{7.18}) and (\ref{7.19}) we find that the density of the
generating functional has the form
\begin{equation}
{\cal F} = - 2i \ln\frac{q Q'}{Q(Q + q)}. \label{7.20}
\end{equation}
Thus we see that canonical symmetry transformation (\ref{7.19}) is a
canonical transformation.

For the transformation of the density of the Hamiltonian we have the
formula
\begin{equation}
{\cal H}(Q, R) = {\cal H}(q, r) + \left( - \frac{4r'q}{1 + rq} +
\frac{2r''}{r'} \right)'. \label{7.21}
\end{equation}

\section{Landau--Lifshits Model}

This model is a generalization of the Heisenberg model. The equations of
motion in this case are
\begin{equation}
\dot{\vec S} = \vec S \times \vec S'' + \vec S \times J \vec S,
\label{8.1}
\end{equation}
where $J$ is a constant symmetric matrix, that can be chosen as a
diagonal one: $J = {\rm diag}(J_1, J_2, J_3)$. Poisson brackets
(\ref{7.3}) lead to equations (\ref{8.1}) if we define the Hamiltonian as
\begin{equation}
H = \frac{1}{2} \int dx\, (\vec S'^2 - \vec S J \vec S). \label{8.2}
\end{equation}
Using the stereographic coordinates, given by (\ref{7.6}), we get the
following expression for the Hamiltonian:
\begin{equation}
H = \int dx\, \left(\frac{2(|\psi'|^2 + \alpha(\psi^2 + \mbar \psi^2) -
\gamma |\psi|^2)}{(1 + |\psi|^2)^2} - \beta \right), \label{8.3}
\end{equation}
where
\begin{equation}
\alpha = \frac{J_2 - J_1}{4}, \qquad \beta = \frac{J_3}{2}, \qquad \gamma
= \frac{J_1 + J_2}{2} - J_3. \label{8.4}
\end{equation}
It is clear that the constant term in the density of the Hamiltonian can
be excluded from the consideration.

The Poisson brackets for $\psi$ and $\mbar\psi$ are given by (\ref{7.8})
and (\ref{7.9}), and the Hamilton equation has the form
\begin{equation}
i \dot \psi + \psi'' - \frac{2 \mbar \psi \psi'^2 - 2 \alpha \psi^3 +
\gamma \mbar \psi \psi^2 - \gamma \psi + 2 \alpha \mbar \psi}{1 +
|\psi|^2} = 0. \label{8.5}
\end{equation}
Denoting
\begin{equation}
\rho(x) = \alpha x^4 + \gamma x^2 + \alpha, \label{8.6}
\end{equation}
we write equation (\ref{8.5}) in a more compact form
\begin{equation}
i \dot \psi + \psi'' - \frac{2 \mbar \psi (\psi'^2 + \rho(\psi))}{1 +
|\psi|^2} + \frac{1}{2}\rho'(\psi) = 0. \label{8.7}
\end{equation}

Construct now a complex extension of the system. The corresponding
equations are
\begin{eqnarray}
&i \dot q + q'' - \frac{2r(q'^2 + \rho(q))}{1 + rq} + \frac{1}{2} \rho'(q)
= 0,& \label{8.8} \\
&i \dot r - r'' + \frac{2q(r'^2 + \rho(r))}{1 + rq} - \frac{1}{2} \rho'(r)
= 0.& \label{8.9}
\end{eqnarray}
These equations are Hamilton equations for the Hamiltonian defined by the
density
\begin{equation}
{\cal H} = \frac{2 (r'q' + \alpha(r^2 + q^2) - \gamma rq)}{(1 + rq)^2}.
\label{8.10}
\end{equation}

The canonical symmetry transformation for the Landau--Lifshits model is
\begin{equation}
Q = \frac{1}{r}, \qquad \frac{1}{1 + RQ} - \frac{1}{1 + rq} = \frac{r r''
- r'^2 + \alpha r^4 - \alpha}{r'^2 + \rho(r)}. \label{8.11}
\end{equation}
It can be shown that this transformation is canonical and the density of
the generating functional has the form
\begin{equation}
{\cal F} = -i \ln \frac{q^2(Q'^2 + \rho(Q))}{Q^2(q + Q)^2} + 2i
\frac{Q'}{\rho^{1/2}(Q)} \arctan \frac{Q'}{\rho^{1/2}(Q)}. \label{8.12}
\end{equation}
Note, that without any loose of generality we can consider $\alpha$ as a
positive constant. In this case the function $\rho(x)$ is positively
definite.

The behaviour of the density of the Hamiltonian under the canonical
symmetry transformation is described by the formula
\begin{equation}
{\cal H}(Q,R) = {\cal H}(q,r) + \left(- \frac{4r'q}{1 + rq} + \frac{2r''r'
+ r'\rho'(r)}{r'^2 + \rho(r)}\right)'. \label{8.13}
\end{equation}

\section{Principal Chiral Field}

The principal chiral field is a function on the space--time taking values
in a Lie group $G$. We consider here the case of the two--dimensional
Minkowski space-time, and use from the beginning a complex extension
of the system, supposing that $G$ is a complex matrix Lie group. The
equations of motion for the principal chiral field $g(t,x)$ have the form
\begin{equation}
(g^{-1}g_{,t})_{,t} - (g^{-1}g_{,x})_{,x} = 0. \label{9.1}
\end{equation}
{}From these equations we see that there exists a function $\vp(t,x)$
taking values in the corresponding Lie algebra $\cal G$ such that
\begin{equation}
g^{-1}g_{,t} = \vp_{,x}, \qquad g^{-1}g_{,x} = \vp_{,t}. \label{9.2}
\end{equation}
If we take some function $\vp$ and solve equations (\ref{9.2}) we
get a solution of equations (\ref{9.1}). Note, that the integrability
conditions of equations (\ref{9.2}), considered as equations for the
function $g$, has the form
\begin{equation}
\vp_{,tt} - \vp_{,xx} = [\vp_{,t}, \vp_{,x}]. \label{9.3}
\end{equation}
Hence, only those functions $\vp$, that satisfy equations (\ref{9.3}),
give solutions of equations (\ref{9.1}). It is clear that any solution
of equations (\ref{9.1}) can be obtained in such a way.  Thus, the
integration problem of equations (\ref{9.1}) is reduced to the integration
problem of equations (\ref{9.3}).

Equations (\ref{9.3}) are Lagrange equations for the action defined by
the density of the Lagrangian
\begin{equation}
{\cal L} = {\rm tr} \left( \frac{1}{2} (\vp_{,t}^2 - \vp_{,x}^2) +
\frac{1}{3} \vp [\vp_{,t}, \vp_{,x}] \right). \label{9.3a}
\end{equation}

We restrict ourself to the case $G = {\rm SL}(2, {\bf C})$, ${\cal G} = {\rm
sl}(2, {\bf C}) = A_1$.  Recall, that the group ${\rm SL}(2, {\bf C})$ is
the group of complex $2\times 2$--matrices with the determinant equal to
one.  The Lie algebra ${\rm sl}(2, {\bf C})$ of the group ${\rm SL}(2, {\bf
C})$ consists of all complex traceless $2\times 2$--matrices. The matrices
\begin{equation}
H = \left(\begin{array}{cr} 1 & 0 \\ 0 & -1 \end{array}\right), \quad
X_+ = \left(\begin{array}{cc} 0 & 1 \\ 0 & 0 \end{array}\right), \quad
X_- = \left(\begin{array}{cc} 0 & 0 \\ 1 & 0 \end{array}\right)
\label{9.4}
\end{equation}
form a basis in the Lie algebra ${\rm sl}(2, {\bf C})$. The commutation
relations for these matrices have the form
\begin{equation}
[H, X_+] = 2 X_+, \qquad [H, X_-] = -2 X_-, \qquad [X_+, X_-] = H.
\label{9.5}
\end{equation}

Represent the field $\vp$ in the form
\begin{equation}
\vp = \vp^- X_- + \vp^0 H + \vp^+ X_+ =
\left(\begin{array}{lr}
\vp^0 & \vp^+ \\
\vp^- & -\vp^0
\end{array}\right). \label{9.6}
\end{equation}
Using this representation, we get from (\ref{9.3}) the following system of
equations for $\vp^-$, $\vp^0$ and $\vp^+$:
\begin{eqnarray}
\vp^-_{,tt} - \vp^-_{,xx} &=& - 2(\vp^0_{,t} \vp^-_{,x} -
\vp^-_{,t} \vp^0_{,x}), \label{9.7} \\
\vp^0_{,tt} - \vp^0_{,xx} &=& \vp^+_{,t} \vp^-_{,x} -
\vp^-_{,t} \vp^+_{,x}, \label{9.8} \\
\vp^+_{,tt} - \vp^+_{,xx} &=& 2(\vp^0_{,t} \vp^+_{,x} -
\vp^+_{,t} \vp^0_{,x}). \label{9.9}
\end{eqnarray}
The density of the Lagrangian can be written for this case in the form
\begin{eqnarray}
{\cal L} = (\vp^0_{,t})^2 &+& \vp^-_{,t} \vp^+_{,t} - (\vp^0_{,x})^2 -
\vp^-_{,x} \vp^+_{,x} + \frac{2}{3} (\vp^0 (\vp^+_{,t} \vp^-_{,x} -
\vp^-_{,t} \vp^+_{,x}) \nonumber \\
&+& \vp^- (\vp^0_{,t} \vp^+_{,x} - \vp^+_{,t} \vp^0_{,x}) - \vp^+
(\vp^0_{,t} \vp^-_{,x} - \vp^-_{,t} \vp^0_{,x})). \label{9.10}
\end{eqnarray}
It is more convenient to use the density of the Lagrangian
\begin{equation}
{\cal L} = (\vp^0_{,t})^2 + \vp^-_{,t} \vp^+_{,t} - (\vp^0_{,x})^2 -
\vp^-_{,x} \vp^+_{,x} + 2 \vp^- (\vp^0_{,t} \vp^+_{,x} -
\vp^+_{,t} \vp^0_{,x}), \label{9.11}
\end{equation}
which differs from the one, given by (\ref{9.10}), by divergency terms.

The canonical symmetry transformation for the case under consideration is
determined from the relations\cite{Lez91a,Lez92}:
\begin{eqnarray}
\Phi^- &=& -\frac{1}{\vp^-}, \label{9.12} \\
\Phi^0_{,t} &=& \vp^0_{,t} - (\vp^0 + \Phi^0) \frac{\vp^-_{,t}}{\vp^-} -
\frac{\vp^-_{,x}}{\vp^-}, \label{9.13} \\
\Phi^0_{,x} &=& \vp^0_{,x} - (\vp^0 + \Phi^0) \frac{\vp^-_{,x}}{\vp^-} -
\frac{\vp^-_{,t}}{\vp^-}, \label{9.14} \\
\Phi^+_{,t} &=& \vp^{-2} \vp^+_{,t} - (\vp^0 + \Phi^0)^2 \vp^-_{,t} -
\vp^-_{,t} \nonumber \\
&&\hspace{4.0em} + 2 \vp^- (\vp^0 + \Phi^0) \vp^0_{,t} - 2
(\vp^0 + \Phi^0) \vp^-_{,x} + 2 \vp^- \vp^0_{,x}, \label{9.15} \\
\Phi^+_{,x} &=& \vp^{-2} \vp^+_{,x} - (\vp^0 + \Phi^0)^2 \vp^-_{,x} -
\vp^-_{,x} \nonumber \\
&&\hspace{4.0em} + 2 \vp^- (\vp^0 + \Phi^0) \vp^0_{,x} - 2 (\vp^0 +
\Phi^0) \vp^-_{,t} + 2 \vp^- \vp^0_{,t}.\hspace{2.0em} \label{9.16}
\end{eqnarray}
In contrast to the examples, considered above, the canonical symmetry
transformation for the principal chiral field is given in an implicit
form. To get the explicit relations we have to integrate
relations (\ref{9.12})--(\ref{9.16}). For example, from (\ref{9.14}) we have
\begin{equation}
\Phi^0 = \partial^{-1} \frac{1}{\vp^-}(\vp^0_{,x} \vp^- - \vp^0 \vp^-_{,x}
- \vp^-_{,t}), \label{9.17}
\end{equation}
where operator $\partial^{-1}$ is defined by (\ref{4.5}).

{}From (\ref{9.11}) we get for the density of the energy of the system the
following expression:
\begin{equation}
{\cal E} = (\vp^0_{,t})^2 + \vp^-_{,t} \vp^+_{,t} + (\vp^0_{,x})^2 +
\vp^-_{,x} \vp^+_{,x}. \label{9.18}
\end{equation}
The direct calculation shows that this density is invariant under the
canonical symmetry transformation. Note, that it is enough here to use
only nonintegrated relations (\ref{9.13})--(\ref{9.16}). From this
invariance it follows that the density of the Hamiltonian of the system is
invariant under the canonical symmetry transformation.

{}From (\ref{9.11}) we see that the generalized momenta are given by
\begin{equation}
\pi^- = \frac{1}{2} \vp^+_{,t}, \qquad \pi^0 = \vp^0_{,t} + \vp^-
\vp^+_{,x}, \qquad \pi^+ = \frac{1}{2} \vp^-_{,t} - \vp^- \vp^0_{,x}.
\label{9.19}
\end{equation}
Consider canonical symmetry transformation (\ref{9.12})--(\ref{9.16}) as a
transformation in the phase space of the system, that connects the
coordinates $\vp^-$, $\vp^0$, $\vp^+$ and $\pi^-$, $\pi^0$, $\pi^+$ with
the new coordinates $\Phi^-$, $\Phi^0$, $\Phi^+$ and $\Pi^-$, $\Pi^0$,
$\Pi^+$. Let us show that this transformation is canonical. It appears
that the quantities $\Phi^-$, $\Phi^0$, $\Phi^+$ and $\pi^-$, $\pi^0$,
$\pi^+$ can be considered as coordinates on the phase space. To
demonstrate this we write the relations
\begin{eqnarray}
\vp^- &=& - \frac{1}{\Phi^-}, \label{9.20} \\
\vp^0\hspace{0.15em} &=& - \Phi^0 + 2 \Phi^- \partial^{-1} \pi^+,
\label{9.21} \\
\vp^+ &=& - \partial^{-1} (\Phi^{-2} \Phi^+_{,x} +\Phi^-_{,x} + 2 \Phi^-
\pi^0 \nonumber \\
&&\hspace{5.0em}{} + 4 \Phi^0_{,x} \Phi^{-2} \partial^{-1} \pi^+ - 4
\Phi^-_{,x} \Phi^{-2} (\partial^{-1} \pi^+)^2), \label{9.22} \\
\Pi^- &=& \frac{\pi^-}{\Phi^{-2}} - \pi^+ + 2 \Phi^- \Phi^+_{,x}
\partial^{-1} \pi^+ + 2 \pi^0 \partial^{-1} \pi^+ \nonumber \\
&&\hspace{5.0em}{} +  \Phi^- \Phi^0_{,x} (\partial^{-1} \pi^+)^2 + 4
\Phi^{-2} \pi^+ (\partial^{-1} \pi^+)^2, \hspace{2.0em} \label{9.23} \\
\Pi^0\hspace{0.2em} &=& - \pi^0 - 2 (\Phi^{-2}(\partial^{-1}
\pi^+)^2)_{,x}, \label{9.24} \\
\Pi^+ &=& - (\Phi^{-2} \partial^{-1} \pi^+)_{,x}. \label{9.25}
\end{eqnarray}
Following the consideration of section 2, we conclude that if the
transformation under consideration is canonical then we can represent
(\ref{9.20})--(\ref{9.25}) in the form
\begin{eqnarray}
&\vp^- = \frac{\delta F}{\delta \pi^-}, \qquad \vp^0 = \frac{\delta
F}{\delta \pi^0}, \qquad \vp^+ = \frac{\delta F}{\delta \pi^+},&
\label{9.26} \\
&\Pi^- = \frac{\delta F}{\delta \Phi^-}, \qquad \Pi^0 = \frac{\delta
F}{\delta \Phi^0}, \qquad \Pi^+ = \frac{\delta F}{\delta \Phi^+},
\label{9.27}
\end{eqnarray}
where $F$ is a functional of $\Phi^-$, $\Phi^0$, $\Phi^+$ and $\pi^-$,
$\pi^0$, $\pi^+$. And vice versa, if we can represent
(\ref{9.20})--(\ref{9.25}) in form (\ref{9.26}), (\ref{9.27}), then we
are dealing with a canonical transformation. It is easy to get convinced,
that if we choose $F$ as the functional, defined by the density
\begin{eqnarray}
{\cal F} &=& {} - \frac{2\pi^-}{\Phi^-} - 2\Phi^0 \pi^0 - 2\Phi^- \pi^+ +
2\Phi^{-2} \Phi^+_{,x} \partial^{-1} \pi^+ \nonumber \\
&&{} + 4 \Phi^- \pi^0 \partial^{-1} \pi^+ + 4 \Phi^{-2} \Phi^0_{,x}
(\partial^{-1} \pi^+)^2 + \frac{8}{3} \Phi^{-3} \pi^+ (\partial^{-1}
\pi^+)^2, \hspace{2.0em} \label{9.28}
\end{eqnarray}
then relations (\ref{9.20})--(\ref{9.25}) will be satisfied. This ends
the proof of the fact, that the canonical symmetry transformation for the
principal chiral field is canonical.

\section{Self--Dual Yang--Mills System}

In this section we consider the self--dual configurations of the
Yang--Mills field in four--dimensional Euclidean space with coordinates
$x^\mu$, $\mu = 1,2,3,4$. We also use the notations $x^1 = t$, $x^2 = u$,
$x^3 = v$ and $x^4 = w$.  Consider the case, when the gauge group $G$
is a complex matrix Lie group.  The gauge potential $a_\mu$ takes values
in the corresponding Lie algebra ${\cal G}$. The field strength
$f_{\mu\nu}$ is defined by
\begin{equation}
f_{\mu\nu} = a_{\nu,\mu} - a_{\mu,\nu} + [a_{\mu}, a_{\nu}]. \label{12.1}
\end{equation}
The self--duality equations have the form
\begin{equation}
f_{\mu\nu} = \frac{1}{2} \varepsilon_{\mu\nu\rho\sigma}f_{\rho\sigma}.
\label{12.2}
\end{equation}
In fact, we have only three independent equations, that can be written as
\begin{equation}
f_{tu} = f_{vw}, \quad f_{tv} = - f_{uw}, \quad f_{tw} = f_{uv}.
\label{12.3}
\end{equation}

Introduce four new complex variables defined by
\begin{equation}
y = t + iu, \qquad \mbar y = t - iu, \qquad
z = v - iw, \qquad \mbar z = v + iw. \label{12.5}
\end{equation}
The components of the field strength in the old and new coordinates are
connected by the relations
\begin{eqnarray}
&f_{y \mbar y} = \frac{i}{2} f_{tu}, \qquad f_{z \mbar z} = -\frac{i}{2}
f_{vw},& \label{12.6} \\
&f_{yz} = \frac{1}{4} (f_{tv} + if_{tw} + f_{uw} - if_{uv}), \quad
f_{\mbar y\mbar z} = \frac{1}{4} (f_{tv} - if_{tw} + f_{uw} + if_{uv}),&
\label{12.10} \\
&f_{y\mbar z} = \frac{1}{4} (f_{tv} - if_{tw} - f_{uw} - if_{uv}), \quad
f_{\mbar y z} = \frac{1}{4} (f_{tv} + if_{tw} - f_{uw} + if_{uv}).&
\hspace{2.0em} \label{12.9}
\end{eqnarray}
It is easy to check that in the new coordinates the self--duality
equations reduce to
\begin{eqnarray}
&f_{yz} = 0, \qquad f_{\mbar y \mbar z} = 0,& \label{12.11} \\
&f_{y\mbar y} + f_{z\mbar z} = 0.& \label{12.12}
\end{eqnarray}
{}From (\ref{12.11}) it follows that we can find two matrix functions $h$
and $\mbar h$ such that
\begin{eqnarray}
&a_y = h^{-1}h_{,y}, \qquad a_z = h^{-1}h_{,z},& \label{12.13} \\
&a_{\mbar y} = \mbar h^{-1}\mbar h_{,\mbar y}, \qquad a_{\mbar z} = \mbar
h^{-1}\mbar h_{, \mbar z}.& \label{12.14}
\end{eqnarray}
Using (\ref{12.13}) and (\ref{12.14}), we get for the field strength
components $f_{y\mbar y}$ and $f_{z \mbar z}$ the representation
\begin{equation}
f_{y \mbar y} = - h^{-1}(g^{-1} g_{,\mbar y})_{,y}h, \qquad f_{z \mbar z} =
- h^{-1}(g^{-1} g_{,\mbar z})_{,z}h, \label{12.15}
\end{equation}
where $g = \mbar h h^{-1}$. Thus, self--duality equation (\ref{12.12})
takes the form
\begin{equation}
(g^{-1} g_{,\mbar y})_{,y} + (g^{-1} g_{,\mbar z})_{,z} = 0. \label{12.16}
\end{equation}
This is the so--called Yang equation.\cite{Yan77,BFNY78}

{}From (\ref{12.16}) we conclude that there exists a function $\vp$,
taking values in the Lie algebra ${\cal G}$, such that
\begin{equation}
g^{-1} g_{,\mbar y} = \vp_{,z}, \qquad g^{-1} g_{,\mbar z} =
-\vp_{,y}. \label{12.17}
\end{equation}
If we chose some function $\vp$, and solve the system of equations
(\ref{12.17}) we get a solution of the Yang equation, and any solution of
this equation can be obtained in such a way. Note, that not every function
$\vp$ can be used in (\ref{12.17}). The integrability conditions of
this system of equation have the form
\begin{equation}
\vp_{,y \bar y} + \vp_{,z \bar z} = [\vp_{,y}, \vp_{,z}],\label{12.18}
\end{equation}
and if we wish to get a solution of the Yang equation we must choose the
function $\vp$, satisfying equation (\ref{12.18}). Thus we reduce the
problem of integration of the Yang equation to the integration of equation
(\ref{12.18}).

Equations (\ref{12.18}) can be obtained from the action
\begin{equation}
S = \frac{1}{2} \int d^4 x\,({\cal L} + \mbar{\cal L}),
\label{12.19}
\end{equation}
where
\begin{equation}
{\cal L} = 2\,{\rm tr}\left(\vp_{,y} \vp_{,\mbar y} + \vp_{,z}
\vp_{,\mbar z} + \frac{2}{3} \vp [\vp_{,y},
\vp_{,z}]\right). \label{12.20}
\end{equation}
The existence of this simple Lagrangian density is the reason for using
equations (\ref{12.18}) instead of the Yang equation.

Consider the case of the Lie algebra ${\rm sl}(2, {\bf C}) = A_1$. In this
case we present $\vp$ in form (\ref{9.6}) and get the following system
of equations
\begin{eqnarray}
\vp^-_{,y \bar y} + \vp^-_{,z \bar z} &=& - 2 (\vp^0_{,y}
\vp^-_{,z} - \vp^-_{,y} \vp^0_{,z}), \label{12.22} \\
\vp^0_{,y \bar y} + \vp^0_{,z \bar z} &=& \vp^+_{,y}
\vp^-_{,z} - \vp^-_{,y} \vp^+_{,z}, \label{12.23} \\
\vp^+_{,y \bar y} + \vp^+_{,z \bar z} &=& 2 (\vp^0_{,y}
\vp^+_{,z} - \vp^+_{,y} \vp^0_{,z}). \label{12.24}
\end{eqnarray}
The density of the Lagrangian up to divergency terms is
\begin{eqnarray}
{\cal L} = 4 (\vp^0_{,\mbar y} \vp^0_{,y} &+& \vp^0_{,\mbar z}
\vp^0_{,z}) + 2 \vp^+_{,\mbar y} \vp^-_{,y} + 2
\vp^-_{,\mbar y} \vp^+_{,y} \nonumber \\
&+& 2 \vp^+_{,\mbar z} \vp^-_{,z} + 2 \vp^-_{,\mbar z}
\vp^+_{,z} + 8\vp^- (\vp^0_{,y} \vp^+_{,z} -
\vp^+_{,y} \vp^0_{,z}). \label{12.25}
\end{eqnarray}

The canonical symmetry transformation for this system has the
form\cite{Lez91b,Lez92}
\begin{eqnarray}
\Phi^- &=& -\frac{1}{\vp^-}, \label{12.26} \\
\Phi^0_{,y} &=& \vp^0_{,y} - (\vp^0 + \Phi^0)
\frac{\vp^-_{,y}}{\vp^-} + \frac{\vp^-_{,\mbar z}}{\vp^-},
\label{12.27} \\
\Phi^0_{,z} &=& \vp^0_{,z} - (\vp^0 + \Phi^0)
\frac{\vp^-_{,z}}{\vp^-} - \frac{\vp^-_{,\bar y}}{\vp^-},
\label{12.28} \\
\Phi^+_{,y} &=& (\vp^-)^2 \vp^+_{,y} + \vp^- [(\vp^0 + \Phi^0)
(\vp^0 + \Phi^0)_{,y} - (\vp^0 + \Phi^0)_{,\mbar z}], \\
\Phi^+_{,z} &=& (\vp^-)^2 \vp^+_{,z} + \vp^- [(\vp^0 + \Phi^0)
(\vp^0 + \Phi^0)_{,z} + (\vp^0 + \Phi^0)_{,\mbar y}]. \label{12.29}
\end{eqnarray}
As in the case of the principal chiral field, the canonical symmetry
transformation is given in an implicit form and require integration.

Using the relations
\begin{equation}
\frac{\partial}{\partial y} = \frac{1}{2}\left( \frac{\partial}{\partial
t} - i \frac{\partial}{\partial u} \right), \qquad
\frac{\partial}{\partial \mbar y} = \frac{1}{2}
\left(\frac{\partial}{\partial t} + i \frac{\partial}{\partial u} \right),
\label{12.30}
\end{equation}
we can rewrite the density of the Lagrangian in the form
\begin{eqnarray}
{\cal L} &=& (\vp^0_{,t})^2 + (\vp^0_{,u})^2 + \vp^+_{,t} \vp^-_{,t} +
\vp^+_{,u} \vp^-_{,u} + 4 \vp^0_{,z} \vp^0_{,\mbar z} + 2 \vp^+_{,\mbar z}
\vp^-_{,z} \nonumber \\
&&{} + 2 \vp^-_{,\mbar z} \vp^+_{,z} + 4 \vp^- (\vp^0_{,t}
\vp^+_{,z} - \vp^+_{,t} \vp^0_{,z} - i \vp^0_{,u}
\vp^+_{,z} + i \vp^+_{,u} \vp^0_{,z}). \label{12.31}
\end{eqnarray}
{}From this formula we get the following expressions for the generalized
momenta:
\begin{equation}
\pi^- = \frac{1}{2} \vp^+_{,t}, \qquad \pi^0 = \vp^0_{,t} + 2 \vp^-
\vp^+_{,z}, \qquad \pi^+ = \frac{1}{2} \vp^-_{,t} - 2 \vp^- \vp^0_{,z}.
\label{12.32}
\end{equation}
It is convenient for our purposes to use instead of the canonical
coordinates $\vp^-$, $\vp^0$, $\vp^+$ and $\pi^-$, $\pi^0$, $\pi^+$
another set of the canonical coordinates $\vp^-$, $\vp^0$, $\vp^+$ and
$\widetilde \pi^-$, $\pi^0$, $\widetilde \pi^+$, where
\begin{equation}
\widetilde \pi^- = \pi^- - \frac{i}{2} \vp^+_{,u}, \qquad \widetilde \pi^+
= \pi^+ + \frac{i}{2} \vp^-_{,u}. \label{12.35}
\end{equation}

To show that the Hamiltonian version of canonical symmetry transformation
(\ref{12.26})--(\ref{12.29}) is a canonical transformation, we look for
the generating functional of the type considered in the previous section.
In other words, we suppose first that the quantities $\Phi^-$, $\Phi^0$,
$\Phi^+$ and $\widetilde \pi^-$, $\pi^0$, $\widetilde \pi^+$ can be
considered as coordinates on the phase space. It is really so, because we
can rewrite relations (\ref{12.26})--(\ref{12.29}) as the relations,
connecting $\vp^-$, $\vp^0$, $\vp^+$ and $\widetilde \Pi^-$, $\Pi^0$,
$\widetilde \Pi^+$ with $\Phi^-$, $\Phi^0$, $\Phi^+$ and $\widetilde
\pi^-$, $\pi^0$, $\widetilde \pi^+$:
\begin{eqnarray}
\vp^- &=& - \frac{1}{\Phi^-}, \label{12.36} \\
\vp^0\hspace{0.15em} &=& - \Phi^0 + \Phi^- \partial^{-1}_{,z} \widetilde
\pi^+, \label{12.37} \\
\vp^+ &=& \partial^{-1}_z ( - \Phi^{-2} \Phi^+_{,z} + \Phi^{-2}
\Phi^-_{,z} (\partial^{-1}_z \widetilde \pi^+)^2 - \Phi^- \pi^0 \nonumber \\
&&{} - 2 \Phi^{-2} \Phi^0_{,z} \partial^{-1}_z \widetilde \pi^+ + i \Phi^-
\Phi^-_{,u} \partial^{-1}_z \widetilde \pi^+ + \Phi^-_{,\mbar z} - i
\Phi^- \Phi^0_{,u}), \label{12.38} \\
\widetilde \Pi^- &=& \frac{\widetilde \pi^-}{\Phi^{-2}} + \pi^0
\partial^{-1}_z \widetilde \pi^+ + \Phi^{-2} \widetilde \pi^+
(\partial^{-1}_z \widetilde \pi^+)^2 + 2 \Phi^- \Phi^0_{,z} (\partial^{-1}
\widetilde \pi^+)^2 \nonumber \\
&&{} + i \Phi^- \partial^{-1}_z \widetilde \pi^+_{,u} \partial^{-1}_z
\widetilde \pi^+ + 2 \Phi^- \Phi^+_{,z} \partial^{-1}_z \widetilde \pi^+ +
i \Phi^0_{,u} \partial^{-1}_z \widetilde \pi^+ + \partial^{-1}
\pi^+_{,\mbar z}, \hspace{2.0em} \label{12.39} \\
\Pi^0 \hspace{0.2em} &=& - \pi^0 - (\Phi^{-2} (\partial^{-1}_z \widetilde
\pi^+)^2)_{,z} - i (\Phi^- \partial^{-1}_z \widetilde \pi^+)_{,u},
\label{12.40} \\
\widetilde \Pi^+ &=& - (\Phi^{-2} \partial^{-1}_z \widetilde \pi^+)_{,z}.
\label{12.41}
\end{eqnarray}
Then we look for the functional $F$ of the variables $\Phi^-$, $\Phi^0$,
$\Phi^+$ and $\widetilde \pi^-$, $\pi^0$, $\widetilde \pi^+$, such that
\begin{eqnarray}
&\vp^- = \frac{\delta F}{\delta \widetilde \pi^-}, \qquad \vp^0 =
\frac{\delta F}{\delta \pi^0}, \qquad \vp^+ = \frac{\delta F}{\delta
\widetilde \pi^+},& \label{12.42} \\
&\widetilde \Pi^- = \frac{\delta F}{\delta \Phi^-}, \qquad \Pi^0 =
\frac{\delta F}{\delta \Phi^0}, \qquad \widetilde \Pi^+ = \frac{\delta
F}{\delta \Phi^+}.& \label{12.43}
\end{eqnarray}
It is easy to show that the functional $F$ does exist and its density has
the form
\begin{eqnarray}
{\cal F} &=& - \frac{2\widetilde \pi^-}{\Phi^-} - 2 \Phi^0 \pi^0 + 2\Phi^-
\partial ^{-1}_z \widetilde \pi^+_{,\mbar z} \nonumber \\
&&{} + 2\Phi^{-2} \Phi^+_{,z} \partial^{-1}_z \widetilde \pi^+ + 2 \Phi^-
\pi^0 \partial^{-1} \widetilde \pi^+ + 2 \Phi^{-2} \Phi^0_{,z}
(\partial^{-1} \widetilde \pi^+)^2 \nonumber \\
&&{} + \frac{2}{3} \Phi^{-3} \widetilde \pi^+  (\partial^{-1}_z \widetilde
\pi^+)^2 + 2 i \Phi^- \Phi^0_{,u} \partial^{-1}_z \widetilde \pi^+ + i
\Phi^{-2} \partial^{-1}_z \widetilde \pi^+_{,u} \partial^{-1}_z \widetilde
\pi^+. \hspace{2.0em} \label{12.44}
\end{eqnarray}
Thus, we see that we again have a canonical transformation.

\section{Conclusion}

The main result of the present paper, from our point of view, is not the
concrete formulae, but the conclusions we can derive from them.

We believe that the integrability of a system is closely related to the
existence of the corresponding canonical symmetry transformation.
Moreover, all the main properties of the system can be derived from the
properties of this transformation. The fact that for all considered cases,
and for many others, which were not included in the present paper, the
canonical symmetry transformations are canonical, is undoubtedly very
important.  But we do not know yet the direct connection of this fact with
the integrability property.

It is known, that the integrability property is connected to the existence
of a nontrivial algebra of internal symmetries. It is possible that this
algebra is hidden in the corresponding canonical symmetry transformation.

\hbox to \hsize{\hfil Received October 23, 1992.}

\end{document}